\documentclass[aps,prb,reprint,superscriptaddress]{revtex4-1}

\usepackage{graphicx}

\begin{document}

\title{Magnetic structure of Cd-doped CeIrIn$_5$}

\author{K.~Beauvois}
\affiliation{Institut Laue Langevin, CS 20156, 38042, Grenoble Cedex 9, France}

\author{N.~Qureshi}
\affiliation{Institut Laue Langevin, CS 20156, 38042, Grenoble Cedex 9, France}

\author{R.~Tsunoda}
\affiliation{Graduate School of Science and Technology, Niigata University, Niigata 950-2181, Japan}

\author{Y.~Hirose}
\affiliation{Department of Physics, Niigata University, Niigata 950-2181, Japan}

\author{R.~Settai}
\affiliation{Department of Physics, Niigata University, Niigata 950-2181, Japan}

\author{D. Aoki}
\affiliation{Institute for Materials Research, Tohoku University, Oarai, Ibaraki, 311-1313, Japan}

\author{P.~Rodi\`{e}re}
\affiliation{Institut N\'{e}el, Universit\'{e} Grenoble Alpes, F-38000 Grenoble, France}
\affiliation{CNRS, Institut N\'{e}el, F-38000 Grenoble, France}

\author{A. McCollam}
\affiliation{High Field Magnet Laboratory (HFML-EMFL), Radboud University, 6525 ED Nijmegen, The Netherlands}

\author{I.~Sheikin}
\email[]{ilya.sheikin@lncmi.cnrs.fr}
\affiliation{Laboratoire National des Champs Magn\'{e}tiques Intenses (LNCMI-EMFL), CNRS, UGA, F-38042 Grenoble, France}

\date{\today}

\begin{abstract}
We report the magnetic structure of nominally 10\% Cd-doped CeIrIn$_5$, CeIr(In$_{0.9}$Cd$_{0.1}$)$_5$, determined by elastic neutron scattering. Magnetic intensity was observed only at the ordering wave vector $Q_{AF} = (1/2,1/2,1/2)$, commensurate with the crystal lattice. A staggered moment of 0.47(3)$\mu_B$ at 1.8~K resides on the Ce ion. The magnetic moments are found to be aligned along the crystallographic $c$ axis. This is further confirmed by magnetic susceptibility data, which suggest the $c$ axis to be the easy magnetic axis. The determined magnetic structure is strikingly different from the incommensurate antiferromagnetic ordering of the closely related compound CeRhIn$_5$, in which the magnetic moments are antiferromagnetically aligned within the tetragonal basal plane.
\end{abstract}

\maketitle

\section{Introduction}

The interplay between antiferromagnetism and unconventional superconductivity remains one of the key questions in Ce-based heavy-fermion compounds. For over a decade, the Ce$M$In\(_5\) ($M =$ Co, Rh, Ir) heavy-fermion materials have served as prototypes for exploring this issue. These compounds crystallize in the tetragonal HoCoGa\(_5\) structure (space group \(P4/mmm\)). CeIrIn$_5$ and CeCoIn$_5$ show superconductivity at ambient pressure below $T_c =$ 0.4 K~\cite{Petrovic2001} and 2.3 K~\cite{Petrovic2001a}, respectively. In CeRhIn$_5$, which is an antiferromagnet with $T_N =$ 3.8 K at ambient pressure, superconductivity with a maximum $T_c$ of 2.1 K occurs in the vicinity of a pressure-induced quantum critical point at the critical pressure \(P_c \simeq \) 2.4 GPa~\cite{Hegger2000,Shishido2005}.

In all three materials, superconductivity is likely to be induced by magnetic quantum fluctuations, which are strongly enhanced in the vicinity of a quantum critical point~\cite{Thompson2012}. It was shown theoretically that $d$-wave superconductivity can be indeed induced by such magnetic fluctuations, but only if they are near commensurate wave vectors~\cite{Hassan2008}. This idea is supported by neutron scattering measurements of CeCoIn$_5$, which have demonstrated a strong coupling between commensurate magnetic fluctuations and superconductivity~\cite{Stock2008}. It is now widely believed that a commensurate magnetic order is favorable for the formation of superconductivity around a quantum critical point in this family of materials. Indeed, a commensurate magnetic order was observed to either coexist or compete with incommensurate ordering in CeRhIn$_5$ doped with either Ir~\cite{Llobet2005} or Co~\cite{Ohira-Kawamura2007,Yokoyama2008}. Remarkably, in these compounds, commensurate antiferromagnetism emerges in the vicinity of a quantum critical point where superconductivity also appears. Interestingly, in CeRh$_{1-x}$Co$_x$In$_5$ a drastic change of the Fermi surface corresponding to the delocalization of Ce $f$-electrons was observed at $x \simeq$ 0.4 where the magnetic order changes from incommensurate to commensurate and superconductivity suddenly emerges~\cite{Goh2008}. It was also argued that particular areas on the Fermi surface nested by the incommensurate wave vector of CeRhIn$_5$ $Q_{AF} = (1/2, 1/2, 0.297)$ play an important
role in forming the superconducting state in CeCoIn$_5$~\cite{Ohira-Kawamura2007}. Furthermore, in Sn-doped CeRhIn$_5$, a drastic change in the magnetic order and a commensurate antiferromagnetism was observed in the proximity of the quantum critical point~\cite{Raymond2014}, where superconductivity is expected, but has not been observed so far. On the other hand, several neutron diffraction experiments performed in CeRhIn$_5$ under pressure up to 1.7~GPa did not reveal the presence of a commensurate antiferromagnetic order~\cite{Majumdar2002,Llobet2004,Raymond2008}. This pressure, however, is considerably lower than the critical value, \(P_c \simeq \) 2.4 GPa, although a pressure-induced bulk superconductivity is observed above about 1.5 GPa~\cite{Knebel2006}.

\begin{figure}[htb]
\includegraphics[width=8cm]{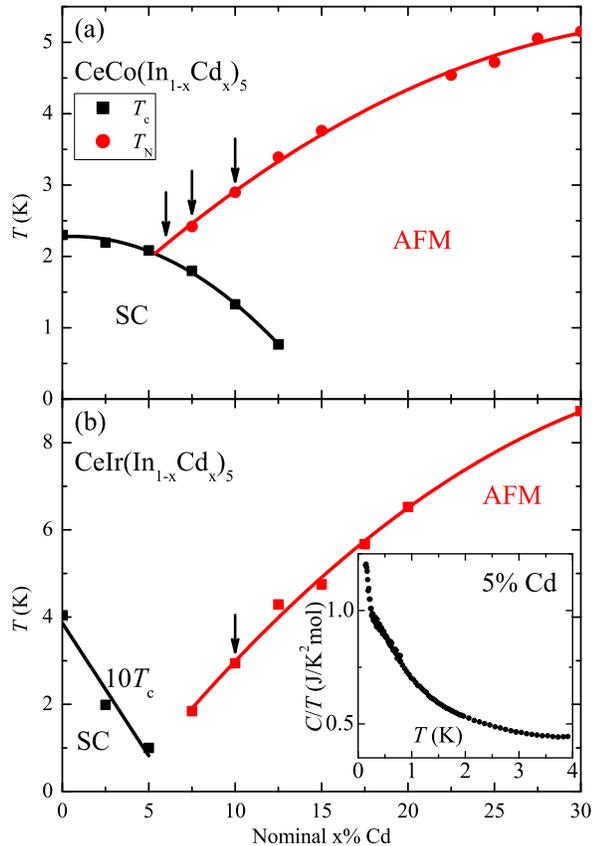}
\caption{\label{Phase_diagram}Doping $x$ dependence of antiferromagnetic (AFM) and superconducting (SC) transition temperatures in (a) CeCo(In$_{1-x}$Cd$_x$)\(_5\) and (b) CeIr(In$_{1-x}$Cd$_x$)\(_5\). All the data are from ref.~\cite{Pham2006}, except for the point for CeIr(In$_{0.95}$Cd$_{0.05}$)$_5$, which is from our own specific heat measurements shown in the inset. Here $x$ is the nominal Cd content of crystals. Arrows indicate Cd concentrations, for which the magnetic structure was determined by neutron diffraction in previous studies for CeCo(In$_{1-x}$Cd$_x$)\(_5\)~\cite{Nicklas2007,Nair2010,Howald2015} and in this work for CeIr(In$_{1-x}$Cd$_x$)\(_5\).}
\end{figure}

In non-magnetic CeCoIn\(_5\) and CeIrIn\(_5\), a quantum critical point can be induced by doping, e.g. by Cd substitution into In sites~\cite{Pham2006}. Temperature-doping phase diagrams obtained from specific heat measurements are shown in Fig.~\ref{Phase_diagram}. Here, $x$ is the nominal concentration of Cd, while the real concentration is about 10 times smaller~\cite{Pham2006}. As shown in Fig.~\ref{Phase_diagram}~(a), introduction of Cd into CeCoIn\(_5\) creates initially a two phase region above nominal $x = 0.075$, where $T_N > T_c$, followed by only antiferromagnetism for $x > 0.12$. The phase diagram of CeIr(In$_{1-x}$Cd$_x$)\(_5\) (Fig.~\ref{Phase_diagram}~(b)) is strikingly different. Only a magnetic ground state is observed beyond the disappearance of superconductivity in the composition range slightly above nominal CeIr(In$_{0.95}$Cd$_{0.05}$)$_5$, for which the superconducting critical temperature is already reduced to about 0.1~K, as shown in the inset of Fig.~\ref{Phase_diagram}~(b).

The magnetic structure of Cd-doped CeCoIn\(_5\) was previously investigated by elastic neutron scattering for nominal Cd concentrations of 6\% ($T_N \approx T_c \approx $ 2 K)~\cite{Howald2015}, 7.5\% ($T_N \approx$ 2.4~K, $T_c \approx$ 1.7 K)~\cite{Nair2010}, and 10\% ($T_N \approx$ 3 K, $T_c \approx$ 1.3 K)~\cite{Nicklas2007}. In all studies, magnetic intensity was observed only at the ordering wave vector $Q_{AF} = (1/2, 1/2, 1/2)$ commensurate with the crystal lattice. This is in line with the above-mentioned hypothesis that a commensurate magnetic order is favorable for heavy-fermion superconductivity, at least in Ce$M$In$_5$ compounds. It would be instructive to further test this hypothesis in Cd-doped CeIrIn\(_5\), for which the magnetic structure is currently unknown.

The above results of the neutron diffraction in single crystals of CeCo(In$_{1-x}$Cd$_x$)\(_5\) are in good agreement with the nuclear-quadrupole-resonance (NQR) measurements in a powder sample of CeCo(In$_{0.9}$Cd$_{0.1}$)\(_5\)~\cite{Yashima2012}. Upon cooling the sample below $T_N$, the NQR line splits into two, indicating a homogenous commensurate antiferromagnetic state with a uniform magnetic moment over the whole sample. On the other hand, in CeIr(In$_{0.925}$Cd$_{0.075}$)\(_5\), the NQR spectrum below $T_N$ exhibits no clear splitting, but a large broadening at its tail, suggesting an inhomogeneous antiferromagnetic order with a large distribution of magnetic moments~\cite{Yashima2012}. The NQR measurements in Cd-doped CeIrIn\(_5\) thus did not shed any light on its magnetic structure, which remains an open question. Therefore, a neutron diffraction experiment is necessary to address this issue.

In this paper, we present neutron diffraction and magnetic susceptibility data on CeIr(In$_{1-x}$Cd$_x$)\(_5\) single crystals with $x = 0.1$, where $x$ represents the nominal concentration, for which the system orders magnetically at $T_N \approx$ 3~K.

\section{Experimental details}

Single crystals of CeIr(In$_{0.9}$Cd$_{0.1}$)$_5$ were grown using a standard In-flux technique with a nominal concentration of 10\% Cd in the indium flux~\cite{Tsunoda2017}. Previous microprobe measurements performed on a series of CeCo(In$_{1-x}$Cd$_{x}$)$_5$ samples grown by the same technique suggest that the actual Cd concentration is only 10\% of the nominal flux concentration~\cite{Pham2006, Nicklas2007}. Although microprobe examination of the CeIr(In$_{1-x}$Cd$_{x}$)$_5$ crystals has not been done, it can be assumed that the Cd concentration in these samples is also approximately 10\% of that in the flux from which they were grown. Therefore, the actual Cd concentration in our sample is likely to be about 1\%.

Neutron diffraction experiment was performed on the D10 beamline of the Institut Laue-Langevin (ILL) of Grenoble (France). For this experiment, we prepared a plateletlike sample with the dimensions $6 \times 5 \times 0.5$~mm$^3$, 0.5~mm being the thickness along the tetragonal $a$ axis, with the other $a$ and $c$ axis being in the plane. The instrument was used in a four-circle configuration with an $80 \times 80$~mm$^2$ two-dimensional microstrip detector. A vertically focusing pyrolytic graphite monochromator was employed, fixing the wavelength of the incoming neutrons to 2.36~\AA. A pyrolytic graphite filter was used in order to suppress higher-order contaminations to 10$^{-4}$ of the primary beam intensity. To reach temperatures down to 1.8~K, we used a closed-cycle cryostat equipped with a Joule-Thompson stage in the four-circle geometry.

The crystal structure was refined using 230 nuclear Bragg peaks. For all peaks, the measured neutron Bragg intensity was corrected for extinction, absorption, and Lorentz factor. For the absorption correction, we accurately modeled the sample shape using the Mag2Pol program~\cite{Qureshi2019}. The obtained lattice parameters at $T =$ 10~K are $a =$ 4.6491(3)~\AA{ }and $c =$ 7.4926(9)~\AA. Regarding the actual Cd concentration and its distribution between the two In sites, it is difficult to obtain a reliable result directly from the refinement using these as adjustable parameters since the scattering lengths of Cd and In are very close to each other. To overcome this issue, we performed the absorption correction with subsequent crystal structure refinement for several Cd concentrations ranging from 1\% to 4\% and assuming equal Cd occupation of the two In sites. The best result was obtained for 3\% Cd concentration. To check the self-consistency of this value, we then used the 3\% Cd absorption corrected data for a refinement with Cd concentration as an adjustable parameter. This yielded the concentration of 0.02(2), consistent with 3\% obtained from the previous iteration. The 3\% Cd concentration estimated in this way in our sample agrees reasonably well with the assumption based on the analogy between In-flux grown samples of CeIr(In$_{1-x}$Cd$_{x}$)$_5$ and CeCo(In$_{1-x}$Cd$_{x}$)$_5$ discussed above.

\section{Results and discussion}

\begin{figure}[htb]
\includegraphics[width=8cm]{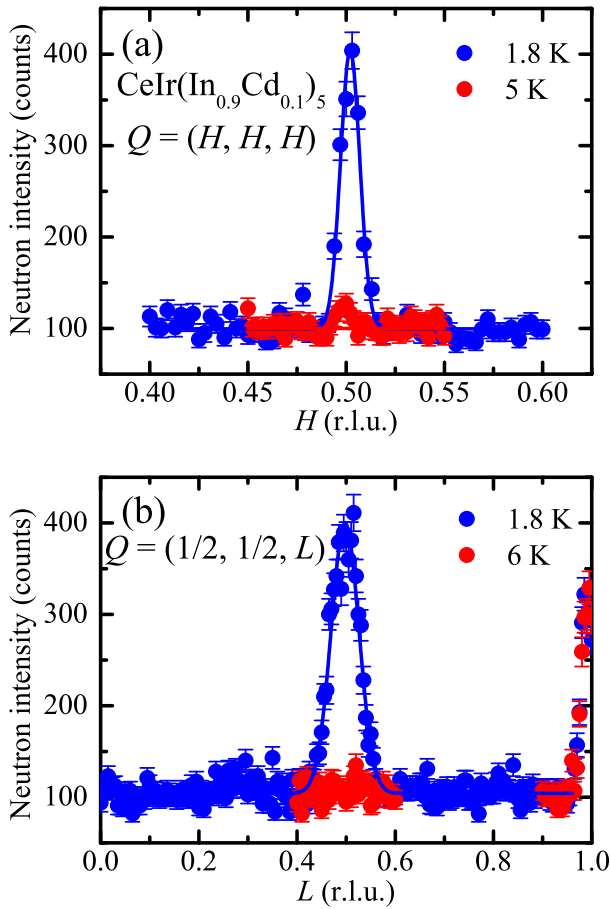}
\caption{\label{Q_scans}Elastic scans in CeIr(In$_{0.9}$Cd$_{0.1}$)$_5$ along the [111] (a) and [001] (b) directions performed below and above the N\'{e}el temperature. Reciprocal lattice units (r.l.u.) are used as coordinates of the reciprocal space. The intensity is in number of counts per $5\times10^5$ monitor counts, which corresponds roughly to 50 s. The solid lines are Gaussian fits of the peaks.}
\end{figure}

Figure~\ref{Q_scans}(a) shows the $\mathbf{Q}$ scan performed along the [111] direction both below (1.8~K) and above (5~K) the N\'{e}el temperature, $T_N$. A clear peak is observed at $\mathbf{Q} =$ (1/2, 1/2, 1/2) at $T =$ 1.8~K, corresponding to a commensurate magnetic wave vector $Q_{AF} =$ (1/2, 1/2, 1/2). The peak disappears at $T =$ 5~K, which is above $T_N$. In order to search for additional peaks with an incommensurate magnetic wave vector, we performed a complete $\mathbf{Q}$ scan in the direction corresponding to the line (1/2, 1/2, $L$) for $0 \leq L \leq 1$ of the reciprocal space at $T =$ 1.8~K, as shown in Fig.~\ref{Q_scans}(b). The latter $\mathbf{Q}$ scan confirmed the presence of a temperature-dependent Bragg peak at $\mathbf{Q} =$ (1/2, 1/2, 1/2). However, we did not observe any additional magnetic Bragg peaks corresponding to an incommensurate magnetic wave vector. In particular, there are no peaks around $\mathbf{Q} =$ (1/2, 1/2, 0.3) characteristic of pure CeRhIn$_5$ at ambient pressure~\cite{Bao2000} or $\mathbf{Q} =$ (1/2, 1/2, 0.4) observed in either doped~\cite{Yokoyama2008,Raymond2014} or pressurized~\cite{Majumdar2002,Raymond2008} CeRhIn$_5$. An increased neutron intensity was observed at $\mathbf{Q} =$ (1/2, 1/2, 1). However, the intensity is temperature-independent, suggesting a $\lambda/2$ contamination from the very strong structural Bragg peak reflection (1, 1, 2) as its origin.

\begin{figure}[htb]
\includegraphics[width=\columnwidth]{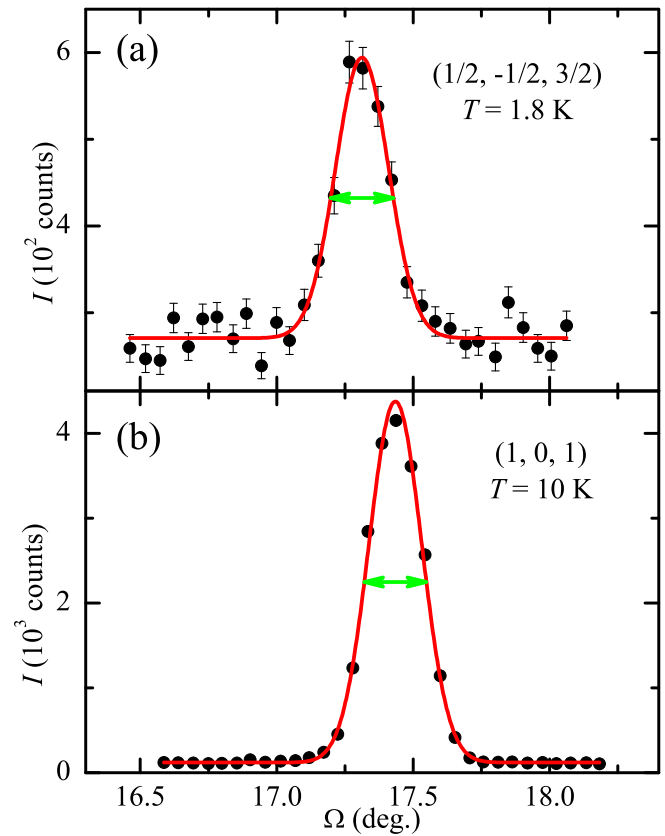}
\caption{\label{Omega_scans}$\Omega$ scans through the magnetic Bragg peak (1/2, -1/2, 3/2) at $T =$ 1.8 K (a) and the nuclear Bragg peak (1, 0, 1) at $T =$ 10 K (b). The intensities are in number of counts per $5\times10^5$ and $3.5\times10^4$ monitor counts for the magnetic and nuclear peaks respectively. Solid lines are Gaussian fits of the peaks. Note that the full width at half maximum (FWHM), indicated by arrows, is the same for the two peaks.}
\end{figure}

The artificial peak at $\mathbf{Q} =$ (1/2, 1/2, 1) is apparently narrower than the magnetic Bragg peak at $\mathbf{Q} =$ (1/2, 1/2, 1/2). In general, broadening of this magnetic peak might be an indication of a slight incommensurability with $Q_{AF} =$ (1/2, 1/2, 1/2$\pm \delta$) or of a reduced correlation length. However, direct comparison of the widths of two peaks even from the same $\mathbf{Q}$ scan is misleading because of different resolution conditions at different values of $\mathbf{Q}$. The proper way to assess the width of a magnetic peak is to compare $\Omega$ scans through both magnetic and a nuclear Bragg peak located at a close 2$\theta$ position. Unfortunately, such a comparison is not possible for the magnetic peak at $\mathbf{Q} =$ (1/2, 1/2, 1/2), as there are no nuclear peaks nearby. On the other hand, this can be conveniently done for the magnetic (1/2, -1/2, 3/2) peak (2$\theta = 34.4^\circ$) and the nuclear (1, 0, 1) peak (2$\theta = 34.6^\circ$). The $\Omega$ scans through these two peaks are shown in Fig.~\ref{Omega_scans}(a) and (b) respectively. The FWHM is 0.23(1)$^\circ$ for the magnetic peak and 0.228(3)$^\circ$ for the nuclear one, i.e. the two peaks have the same width within the error bar. Therefore, our data do not reveal any indication for an incommensurate magnetic order or a reduced correlation length.

\begin{figure}[htb]
\includegraphics[width=8cm]{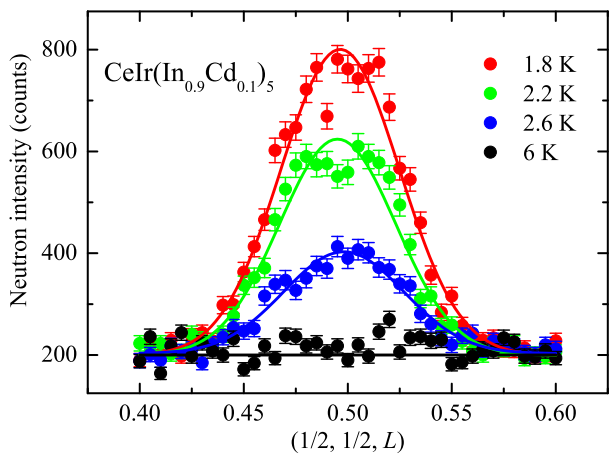}
\caption{\label{Q-scan_T-dependence}$\mathbf{Q}$ scans performed along the [001] direction at different temperatures. The intensity is in number of counts per $1\times10^6$ monitor counts, which corresponds roughly to 100 s. The lines are Gaussian fits of the peaks.}
\end{figure}

\begin{figure}[htb]
\includegraphics[width=8cm]{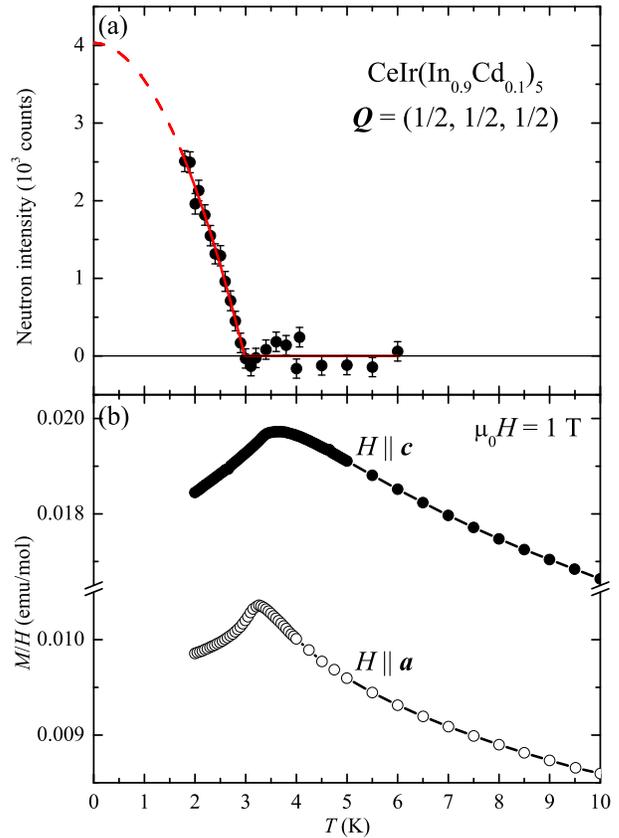}
\caption{\label{T_dep}(a) Temperature dependence of the (1/2, 1/2, 1/2) magnetic Bragg peak intensity  after subtracting the background. The intensity is in number of counts per $4.5\times10^6$ monitor counts, which corresponds roughly to 8 min. The line is a phenomenological fit as explained in the text. (b) Magnetic susceptibility measured in magnetic field of 1~T applied both along the $c$ and $a$ axis as a function of temperature.}
\end{figure}

Figure~\ref{Q-scan_T-dependence} shows elastic scans along the [001] direction across $\mathbf{Q} =$ (1/2, 1/2, 1/2) at different temperatures. The magnetic Bragg peak at $\mathbf{Q} =$ (1/2, 1/2, 1/2) does not shift with temperature; only its intensity decreases with increasing temperature. Since the position of the magnetic propagation vector in reciprocal space does not change with temperature, the temperature dependence of the neutron diffraction intensity at the center of the magnetic Bragg peak $\mathbf{Q}$ was recorded (Fig.~\ref{T_dep}(a)). This intensity, $I$, is proportional to the square of the ordered magnetic moment. To determine the N\'{e}el temperature, the data were fitted by a phenomenological function $I/I_0 = 1 - (T/T_N)^\alpha$, with $\alpha$ a free parameter. This function was successfully used to fit the temperature dependence of the magnetic Bragg peak intensity in other heavy fermion compounds, such as CePd$_2$Si$_2$~\cite{Dijk2000,Kernavanois2005}, Sn-doped CeRhIn$_5$~\cite{Raymond2014}, Cd-doped CeRhIn$_5$~\cite{Howald2015}, and CePt$_2$In$_7$~\cite{Raba2017}. The best fit is obtained with $\alpha = 2.0\pm0.5$ and $T_N = 3.0\pm0.1$~K. The latter value is consistent with $T_N$ determined from previous specific heat~\cite{Pham2006} and resistivity~\cite{Tsunoda2017} measurements. It is also in agreement with the magnetic susceptibility data shown in Fig.~\ref{T_dep}(b).

In order to refine the magnetic structure, a data set of 13 magnetic Bragg reflections was collected at $T =$ 1.8~K. As the magnetic peaks are rather weak, they were integrated using the RPlot program~\footnote{Available at: https://www.ill.eu/users/instruments/instruments-list/d10/software/} for a small detector area around its center to reduce the background noise. The same mask on the detector was used for the integration of nuclear peaks. For all magnetic peaks, the correction for extinction, absorption, and Lorentz factor was performed the same way as for nuclear peaks described above. The resulting intensities are shown in Table~\ref{tab:mag_peaks}. The refinement was then performed using the Mag2Pol program~\cite{Qureshi2019}. Only two arrangements of the magnetic moments are allowed by the group theory: they can be either aligned in the basal plane or along the $c$ axis. For both structures, the calculated intensities are also shown in Table~\ref{tab:mag_peaks}. A much better result is obtained for the $c$ axis configuration, i.e. magnetic moments antiferromagnetically aligned along the $c$ axis. This structure is consistent with the magnetic susceptibility data shown in Fig.~\ref{T_dep}(b), which suggest the $c$ axis to be the easy magnetic axis. The staggered magnetic moment is found to be $M = 0.47(3)\mu_B$ per Ce atom at $T =$ 1.8~K. It is obvious, however, that the magnetic intensity does not saturate at $T =$ 1.8~K (Fig.~\ref{T_dep}(a)). The extrapolation of $I(T)$ to $T = 0$ yields the magnetic moment $M \approx 0.6\mu_B$/Ce at the zero temperature limit. This value is similar to $M \sim 0.7\mu_B$/Ce estimated in CeCo(In$_{0.9}$Cd$_{0.1}$)$_5$ from NQR measurements~\cite{Urbano2007}. Remarkably, both CeIrIn$_5$ and CeCoIn$_5$ doped with 10\% Cd undergo an antiferromagnetic transition at about the same temperature, $T_N \approx$ 3 K (Fig.~\ref{Phase_diagram}).

\begin{table}[hbt]
\caption{\label{tab:mag_peaks}Magnetic refinement for the two possible magnetic structures, as discussed in the text. $I_{\mathrm{calc}}^c$ and $I_{\mathrm{calc}}^{ab}$ are the calculated intensities for magnetic moments aligned along the $c$ axis and in the basal plane respectively. (Note: $R_c$ = 18.7, $R_{ab}$ = 48.1).}
\begin{ruledtabular}
\begin{tabular}{c c c c}
  $\mathbf{Q}$ & $I_{\mathrm{obs}}$ & $I_{\mathrm{calc}}^c$ & $I_{\mathrm{calc}}^{ab}$\\
  \hline
  (1/2, -1/2, 1/2) & 1024(49) & 504 & 205 \\
  (1/2, -1/2, 3/2) & 294(27) & 167 & 178 \\
  (1/2, -1/2, 5/2) & 87(10) & 57 & 131 \\
  (3/2, -1/2, -1/2) & 286(42) & 343 & 133 \\
  (3/2, -1/2, 3/2) & 208(15) & 226 & 116 \\
  (3/2, -3/2, -1/2) & 193(20) & 236 & 91 \\
  (3/2, -3/2, 3/2) & 190(15) & 175 & 79 \\
  (1/2, -1/2, 7/2) & 32(5) & 20 & 84 \\
  (5/2, 1/2, 1/2) & 137(17) & 164 & 35\\
  (5/2, 3/2, 1/2) & 115(25) & 116 & 10\\
  (1/2, 5/2, 3/2) & 111(15) & 128 & 13\\
  (5/2, 3/2, 3/2) & 166(25) & 93 & 11\\
  (5/2, 5/2, 1/2) & 190(77) & 60 & 1\\
\end{tabular}
\end{ruledtabular}
\end{table}

At first glance, it is surprising why the commensurate magnetic order with such a strong staggered magnetic moment was not observed in NQR measurements in Cd-doped CeIrIn$_5$~\cite{Yashima2012}. These measurements, however, were performed on a sample with nominal Cd concentration of 7.5\%, in which the magnetic moment is likely to be reduced with respect to CeIr(In$_{0.9}$Cd$_{0.1}$)$_5$ studied here. Furthermore, in CeIr(In$_{0.925}$Cd$_{0.075}$)\(_5\) the antiferromagnetic transition temperature is lower than 2~K~\cite{Pham2006}. This implies that the NQR measurements, performed at 1.5~K, were done barely below the N\'{e}el temperature, where the ordered moment is presumably very small. Finally, the NQR measurements were carried out on a powder sample, which alone might strongly affect the result. For example, NQR measurements performed on a powder sample of CeRhIn$_5$ under pressure revealed a change of magnetic structure from incommensurate to commensurate~\cite{Yashima2009}, while no such change was observed in either neutron diffraction~\cite{Majumdar2002,Llobet2004,Raymond2008} or single-crystal NQR measurements~\cite{Yashima2020}. From this point of view, it would be interesting to perform NQR measurements on a single crystal of CeIr(In$_{0.9}$Cd$_{0.1}$)$_5$.

The antiferromagnetic arrangement of the magnetic moments, which are aligned along the $c$ axis in CeIr(In$_{0.9}$Cd$_{0.1}$)$_5$, is different from both pure and doped CeRhIn$_5$. Indeed, the ordered moments were found to lie in the tetragonal basal plane in the incommensurate phase of CeRhIn$_5$~\cite{Bao2000,Bao2003,Raymond2007,Fobes2017}, and both in the incommensurate and commensurate states of CeRh$_{1-x}$Ir$_x$In$_5$~\cite{Llobet2005}. On the other hand, the same antiferromagnetic arrangement of the ordered moments along the $c$ axis occurs in U-based 115 compounds, such as UNiGa$_5$~\cite{Tokiwa2002} and, most probably, URhIn$_5$~\cite{Sakai2013}.

The commensurate magnetic order with $Q_{AF} = (1/2, 1/2, 1/2)$ determined here for CeIr(In$_{0.9}$Cd$_{0.1}$)$_5$ is also strikingly different from that in pure CeRhIn$_5$, in which an incommensurate magnetic structure with $Q_{AF} = (1/2, 1/2, 0.297)$ was reported~\cite{Bao2000,Bao2003,Raymond2007,Fobes2017}. On the other hand, the same commensurate magnetic ordering wave vector was observed in Cd-doped CeCoIn$_5$~\cite{Nicklas2007,Nair2010,Howald2015}, in which superconductivity coexists with magnetic order over a wide range of Cd concentrations (Fig.~\ref{Phase_diagram}(a)). Furthermore, the same commensurate magnetic structure with ordering wave vector $(1/2, 1/2, 1/2)$ develops in the doping series CeRh$_{1-x}$Ir$_x$In$_5$~\cite{Llobet2005} and CeRh$_{1-x}$Co$_x$In$_5$~\cite{Ohira-Kawamura2007,Yokoyama2008} at low temperatures over the doping range, for which superconductivity coexists with antiferromagnetism. At first glance, this appears surprising given that superconductivity does not coexist with antiferromagnetic order in Cd-doped CeIrIn$_5$ (Fig.~\ref{Phase_diagram}(b)). A possible clue to this puzzle is offered by In-NQR measurements in Cd-doped CeIrIn$_5$~\cite{Yashima2012}. These measurements suggest that superconductivity in pure and Cd-doped CeIrIn$_5$ is likely mediated by valence fluctuations, and not spin fluctuations, as in other Ce$M$In$_5$ compounds.

\section{Conclusions}

In summary, we carried out elastic neutron scattering experiments on CeIr(In$_{0.9}$Cd$_{0.1}$)$_5$. At low temperatures, we found magnetic intensity at the commensurate wave vector $Q_{AF} = (1/2, 1/2, 1/2)$. The magnetic intensity is building up below $T_N \approx$ 3~K, with $T_N$ being in good agreement with specific heat~\cite{Pham2006}, resistivity~\cite{Tsunoda2017}, and magnetic susceptibility data. No indication for additional intensity was observed at incommensurate positions, such as $(1/2, 1/2, 0.297)$, where CeRhIn$_5$, the related antiferromagnetic member of the Ce$M$In family, orders. A magnetic moment of 0.47(3)$\mu_B$ at 1.8~K resides on the Ce ion, and the moments are antiferromagnetically aligned along the $c$ axis. This is again in contrast to CeRhIn$_5$, in which magnetic moments are antiferromagnetically aligned in the basal plane.

\begin{acknowledgments}
We thank E. Ressouche, B. Ouladdiaf, and C. Simon for fruitful discussions. This work was partially supported by the ANR-DFG grant ``Fermi-NESt''.
\end{acknowledgments}

\bibliography{CeIrIn5Cd}

\end{document}